# Enhancing the Understanding of Computer Networking Courses through Software Tools

Dr. Z. I. Dafalla
College of Applied Sciences – Salalah,
Sultanate of Oman.
zubeir.sal@cas.edu.om

Dr. RD.Balaji,
College of Applied Sciences - Salalah,
Sultanate of Oman,
balajicas@yahoo.com

## ABSTRACT

Computer networking is an important specialization in Information and Communication Technologies. However imparting the right knowledge to students can be a challenging task due to the fact that there is not enough time to deliver lengthy labs during normal lecture hours. Augmenting the use of physical machines with software tools help the students to learn beyond the limited lab sessions within the environment of higher Institutions of learning throughout the world. The Institutions focus mostly on effective use of available resources i.e. a lab may have few lab sessions scheduled within a day for different courses. Hence a particular lab session must begin and end on time. A slow student who did not complete his/her lab exercise must vacate the lab because another class is about to commence. Hence using free software tools such as OPNET IT guru, Packet trace and NS2 will help the student to learn beyond the College time. A Student will gain an insight into Computer networking field by repeatedly doing the same labs at the comfort of his laptop or PC at home, hence increasing deeper understanding into the subject. The main objective of this paper is to explore the software tools which will enhance the network's practical skills among higher education students at College of Applied Sciences Salalah. The paper also discusses the methodologies of implementing and executing lab sessions after the normal college hours. The limitations and suggestions for the improvements are also specified at the end of this paper.

## INTRODUCTION

Education is defined in a different way by dictionaries and academicians. But in broader sense it is viewed in two different perspectives. They are teaching and learning. The learning is nothing but any action or continuous actions performed by a person which make some formative effects on the mind of that person. Mainly education helps to transmit the accumulated knowledge, skills, ethics, values and culture of the society to the next generation deliberately [1].

The traditional teaching differs a lot from the contemporary teaching due to the influence of the contemporary technologies. The technology education still difference from the normal learning process because it needs to learn not only the knowledge but also skills in that technology. Technology education needs the knowledge and skills from the components of science, technology, engineering and mathematics (STEM) [2]. In general technology helps humankind to make their life easy. We do our most of jobs fast and easy because of the advancements in the technology. All technologies are having application of tools, materials, processes and systems done by us to solve our problems. Apart from the normal learning tools like reading books, listening and viewing videos and audio files, technology education need to have practice of few skills by activities in the labs [3]. In the technology learning knowledge of content, processes and skills should be used together to effectively engage students and promote a complete understanding of the sciences, related technologies and their interrelationship.

Nowadays, in all countries technological education is mainly focused to show their strength and development to others. In most of the universities makes the lab courses separately from the other courses to give better training in the technological education. Instructional technology is another stream which mostly concentrate on the delivering the courses to students through ICT. In this paper we are concentrating on delivering the technology education with the help of instructional technology to the higher education students.

**LAB BASED LEARNING**

Technological education cannot be completed without laboratory sessions. In laboratories students can think, discuss and they can try to simulate or really implement what they have learned in the classroom [4].

McKeachie said, "Laboratory teaching assumes that first-hand experience in observation and manipulation of the materials of science is superior to other methods of developing understanding and appreciation. Laboratory training is also frequently used to develop skills necessary for more advanced study or research" about lab based learning. (in Gage, 1962, p. 1144-1145).

The students can try to do what the professionals of that domain can do in the labs. Labs make the students understanding in a specific concept very effective and relation with the real word applications. The interest of the students will be aroused by the lab sessions [5].

Shulman and Tamir, in the Second Handbook of Research on Teaching (Travers, ed., 1973), listed five types of objectives that may be achieved through the use of the laboratory in science classes:

1. Skills - manipulative, inquiry, investigative, organizational, communicative
2. Concepts - for example, hypothesis, theoretical model, taxonomic category
3. Cognitive abilities - critical thinking, problem solving, application, analysis, synthesis
4. Understanding of the nature of science- scientific enterprise, scientists and how they work, existence of a multiplicity of scientific methods, interrelationships between science and technology and among the various disciplines of science

5. Attitudes - for example, curiosity, interest, risk taking, objectivity, precision, confidence, perseverance, satisfaction, responsibility, consensus, collaboration, and liking science (1973, p.1119).

Usually lab sessions are created by the academicians for the students in more skillful and creative way. These lab sessions will be the first research projects of the students in that domain. Students get more indirect skills also along with domain knowledge. It develops critical thinking, data analysis skills and reporting skills through lab session. It also helps the students to see the concepts in a different view. The lab sessions should not disturb the conceptual teachings. Sometimes these lab sessions required more time for some students based on the speed of their learning capability. Hence lecturers need to balance between different categories of students during the lab sessions. College of Applied Sciences Salalah wants to practice these lab sessions with more effective way without disturbing the normal class, as well as helps the students to learn on their own pace. These concepts are discussed in the following sections.

**FLIPPED CLASSROOM**

Flipped Classrooms are a boon for academicians, which helps to utilize the students time even after their class hour for teaching using various technologies like QR code or E-books [7][8]. " The flipped classroom inverts traditional teaching methods, delivering instruction online outside of class and moving "homework" in to the class room" [6]. Flip class room helps the slow learners to learn their courses at their own pace. Contemporary technologies help learners to communicate with their peer or their lecturers with the help of social network, forums, live chat, etc.,. The face to face interaction facility in the traditional teaching replaces by these technologies and make the students to learn comfortably even outside the classroom. In the flipped classroom the main concept used is the activity based learning [6]. Since 2007 this methodology used by the academicians and now it is also used by the MOOCs. Many researchers proved that the flipped classroom helps to improve the understanding of the students in a specific topic and give the better result in overall [7] [8]. Flipped classrooms are very much helpful to the students to revisit the lectures, since the lectures are small in size (8 to 15 minutes), when they don't understand the concept. We are trying to use the same concept in the College of Applied Sciences Salalah for the network courses so that students will get more time for the lab practices and also they don't lose time for the conceptual topics.

**REFERENCES**

[1] Atherton J S (2013) Learning and Teaching; What is learning? retrieved 11 January 2015 from http://www.learningandteaching.info/learning/whatlearn.htm

[2] Technology Education, retrieved 24 December 2014 from http://www.education.state.pa.us/portal/server.pt/community/technology_education/14635

[3] Lab Education, retrieved 26 December 2014 from http://d-lab.mit.edu/courses/education

[4] Gopal, T., Herron, S. S., Mohn, R. S., Hartsell, T., Jawor, J. M., & Blickenstaff, J. S. (2010). Effect of an interactive web-based instruction in the performance of undergraduate anatomy and physiology lab students. Computers & Education, 55, 500 - 512. http://dx.doi.org/10.1016/j.compedu.2010.02.013

[5] Schaefer, D., Scott, D.W., Molina, G. J., Al-Kalaani, Y., Murphy, T., Johnson, W., & Thamburaj Goeser, P. (2008). Integration of Distance Learning Technology into Traditional Engineering Physical Laboratory Exercises. ASEE Southeast Section Conference. http://www.srl.gatech.edu/Members/dschaefer/Publications/Final.RP2008044SCH.pdf

[6] "The Flipped Classroom: Turning Traditional Education On Its Head", retrieved 11 November 2014 from http://www.knewton.com/flipped-classroom/

[7] Dr. RD.Balaji, et al, 2014, "Effectiveness of QR Code in Educational ICT", 2nd ICAICT 2014, 28th & 29th April, 2014, Middle East College, Muscat, Oman.

[8] Dr. RD.Balaji, et al, 2013, "QR With Moodle for Effective Higher Education", IJRCM, Volume No 3, Issues No 4, pp. 14-17, ISSN 2231 1009, April 2013